\documentclass[twocolumn,amsmath,prb,superscriptaddress,nofootinbib]{revtex4-1}

\pdfoutput=1
\usepackage{upgreek}
\usepackage{bm}
\usepackage{pifont}
\usepackage{graphicx}
\usepackage{color}
\usepackage[colorlinks=true, pdfstartview=FitV, linkcolor=blue, citecolor=blue, urlcolor=blue]{hyperref}
\newcommand{\assign}{:=}
\newcommand{\tmmathbf}[1]{\ensuremath{\boldsymbol{#1}}}

\begin{document}
\preprint{APS/123-QED}

\title{Spin Waves and Dirac Magnons in a Honeycomb Lattice Zig-zag Antiferromagnet BaNi$_2$(AsO$_4$)$_2$}

\author{Bin Gao}
\thanks{These two authors contributed equally}
\author{Tong Chen}
\thanks{These two authors contributed equally}	
\affiliation{Department of Physics and Astronomy,
Rice University, Houston, Texas 77005, USA}
\author{Chong Wang}
\affiliation{Department of Physics, Carnegie Mellon University, Pittsburgh, Pennsylvania 15213, USA}
\author{Lebing Chen}
\affiliation{Department of Physics and Astronomy,
Rice University, Houston, Texas 77005, USA}
\author{Ruidan Zhong}
\affiliation{Tsung-Dao Lee Institute and School of Physics and Astronomy, Shanghai Jiao Tong University, Shanghai 200240, China}
\author{Douglas L. Abernathy}
\affiliation{Neutron Scattering Division, Oak Ridge National Laboratory, Oak Ridge, Tennessee 37831, USA}
\author{Di Xiao}
\email{dixiao@andrew.cmu.edu}
\affiliation{Department of Physics, Carnegie Mellon University, Pittsburgh, Pennsylvania 15213, USA}
\author{Pengcheng Dai}
\email{pdai@rice.edu}
\affiliation{Department of Physics and Astronomy,
Rice University, Houston, Texas 77005, USA}

\date{\today}

\begin{abstract}
The topological properties of massive and massless fermionic    
quasiparticles have been intensively investigated over the past decade in topological materials without magnetism. 
Recently, the bosonic analogs of such quasiparticles arising from spin waves 
have been reported in the two-dimensional (2D) honeycomb lattice ferromagnet/antiferromagnet and  
the 3D antiferromagnet. 
Here we use time-of-flight inelastic neutron scattering to study spin waves of the $S=1$ honeycomb lattice antiferromagnet BaNi$_2$(AsO$_4$)$_2$, which has a zig-zag antiferromagnetic (AF) ground state identical to that of the 
Kitaev quantum spin liquid candidate $\alpha$-RuCl$_3$.  We determine the magnetic exchange interactions 
in the zig-zag AF ordered phase, and show that spin waves in BaNi$_2$(AsO$_4$)$_2$ have 
symmetry-protected Dirac points inside the Brillouin zone boundary. These results provide a microscopic understanding of the 
zig-zag AF order and associated Dirac magnons in honeycomb lattice magnets, and are also important for establishing the magnetic interactions in Kitaev quantum spin liquid candidates.
\end{abstract}

\maketitle
\section{introduction}
Elucidating nontrivial crossing points in the band structure of a crystalline solid plays an important role 
in understanding its the momentum space topology \cite{Hazan2010, Qi2011, Haldane2017}. 
The discovery of massless Dirac fermions in the electron bands of graphene has led to intensive studies of 
topological phases in metallic systems \cite{Zhang2005, Novo2005}.
A plethora of topologically nontrivial electron band structures with massless and massive fermionic quasiparticles has been proposed and studied in electronic systems \cite{Chiu2016, Armitage2018}. However, band topology is not restricted to fermionic systems and can also be 
extended to bosonic quasiparticles \cite{Lu2014, Ozawa2019,Susstrunk2016, Zhang2018,Li2016,Pershoguba2018,Chen2018,Cai2021,Chen2021,FFZhu2021,Yao2018,Bao2018,Yuan2020,Boyko2018,Lee2018}. In magnetic ordered materials where spin waves are bosons, topological bosonic quasiparticles (magnons) 
are characterized by chiral edge states and a 
spin gap at the Dirac points induced by the next-nearest-neighbor Dzyaloshinskii-Moriya (DM) interaction, 
as found in the 2D honeycomb ferromagnet CrI$_3$/CrBr$_3$/CrGeTe$_3$ \cite{Chen2018,Cai2021,Chen2021,FFZhu2021}, 
or symmetry-protected band crossings, 
such as the magnon Dirac cones in the 3D antiferromagnet Cu$_3$TeO$_6$ \cite{Yao2018,Bao2018} and the 2D honeycomb lattice antiferromagnet CoTiO$_3$ \cite{Yuan2020}.  Recently, topological properties of magnon bands in collinear magnetic orders have been investigated on the honeycomb lattice.  While the ferromagnetic (FM) phase exhibits a magnon spectrum similar to the electron dispersion of graphene, the antiferromagnetic (AF) phases show an even richer magnon structure.  In particular,
the zig-zag AFM ordered phase, shown in Figs. 1(a,b), can host a Dirac nodal 
line protected by nonsymmorphic symmetry combined with time-reversal symmetry \cite{Boyko2018, Lee2018}.

In addition to potentially hosting topological magnon bands, 2D honeycomb lattice magnetic materials 
themselves are of great interest because the Kitaev spin model \cite{Kitaev}, 
consisting of a network of spins with $S=1/2$ on a honeycomb lattice, 
is predicted to host quantum spin liquid (QSL) with Majorana fermions as its excitations important for quantum computation \cite{Takagi2019}.
Although there is no confirmed Kitaev QSL material, $\alpha$-RuCl$_3$ has been identified as a candidate with 
a zig-zag AF ground state [Fig. 1(b)] \cite{Cao2016}.  One unsolved quest is to determine magnetic exchange couplings giving rise to the zig-zag AF order, which is necessary to extract Kitaev interactions in $\alpha$-RuCl$_3$ \cite{Rau2016}.
Actually, this quest has not been completely solved in all 2D honeycomb lattice magnets with a zig-zag AF ground state, despite many compounds have been studied \cite{Zvereva2015,Bera2017,Zvereva2016,FYe2012,kim2020,Wong2016,Kurbakov2020}.
Inelastic neutron scattering (INS) experiments carried out on powder samples of some of these compounds provided constraints on the magnetic
exchange couplings \cite{kim2020,Choi2012,Songvilay2020}. Nevertheless, 
no measurements on single crystals have conclusively unveiled the magnetic couplings giving rise to the zig-zag AF structure. 

BaNi$_2$(AsO$_4$)$_2$ is a rare example of 2D honeycomb lattice magnet with $S=1$ and a zig-zag AF structure [Figs. 1(a) and 1(b)] \cite{Regnault1980}, which magnetic exchange couplings could be completely determined. 
Unfortunately, early INS experiments carried out on single crystals of BaNi$_2$(AsO$_4$)$_2$ and related 
compounds \cite{Regnault83,Regnault1986,Regnault1983,Regnault1990} have not mapped out the entire spin wave spectra, due to limitation of the spectrometers used nearly four decades ago.  
In addition, the isostructural compound BaCo$_2$(AsO$_4$)$_2$, 
which has a helical magnetic ground state at zero field \cite{Regnault2018}, 
has been recently proposed as a candidate for a field-induced Kitaev QSL state \cite{Zhong2020}. 
Therefore, it is of great interest to map out spin waves in BaNi$_2$(AsO$_4$)$_2$ and 
determine the magnetic exchange couplings and topological properties. 
\begin{figure}[t]
\centering
\includegraphics[scale=.8]{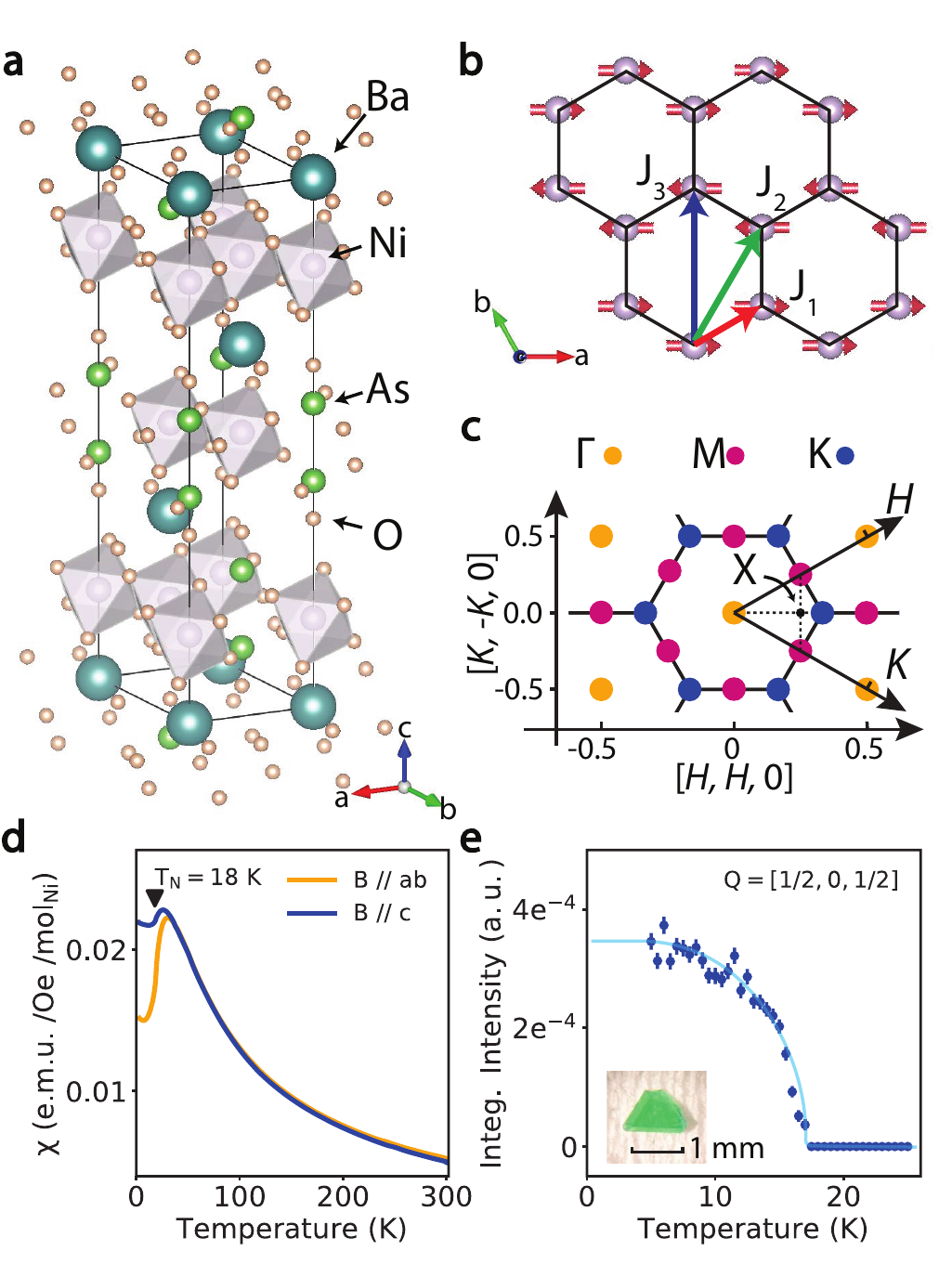}
\caption{\label{fig1}(a) Crystal structure of BaNi$_2$(AsO$_4$)$_2$. (b) Top-down view of the nickel honeycomb lattice with Heisenberg exchange paths. (c) The Brillouin zone boundaries and high symmetry points in the [\emph{H}, \emph{K}] plane. (d) Susceptibility of BaNi$_2$(AsO$_4$)$_2$ measured with 1 T magnetic field applied parallel to the $c$ axis and $ab$ plane. The sharp turn in susceptibility data indicates a magnetic phase transition at $T_N = 18$ K. (e) Order-parameter-like $(1/2, 0, 1/2)$ magnetic Bragg peak. Inset: photo of the single crystal used in the susceptibility measurement.}
\end{figure}

\begin{figure*}[t]
\centering
\includegraphics[scale=.8]{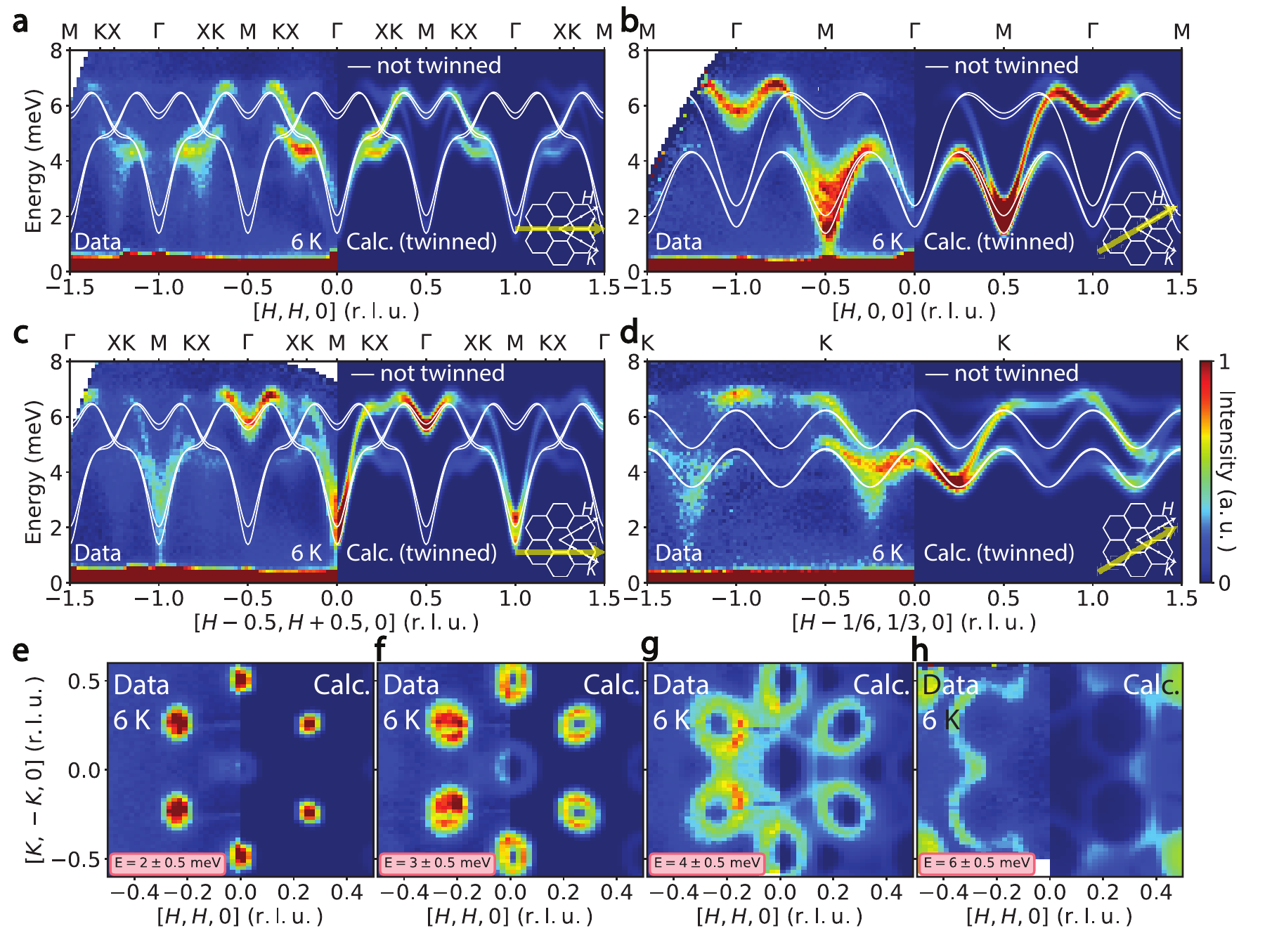}
\caption{\label{fig2} (a-d) Neutron scattering $E$-${\bf Q}$ spectra along high-symmetry directions experimentally at 6 K (left), 
compared with the calculated intensities and dispersion (twinned) discussed in the text (right). 
The line plots of a selected domain (not twinned) were overplotted on the measured and calculated dispersion.
The experimental data were integrated over $-4 \leq L \leq 4$ and $-0.1 \leq K_\perp \leq 0.1$. 
(e-h) Constant-energy cuts at selected energy-transfers in the $[H, K]$ plane at 6 K compared with the calculated pattern. 
The experimental data were integrated over $-4 \leq L \leq 4$.}
\end{figure*}

In this paper, we report INS studies of spin waves on single crystals of BaNi$_2$(AsO$_4$)$_2$. 
Using neutron time-of-flight spectroscopy, 
we map out spin waves in the AF ordered state and fit the data with a Heisenberg Hamiltonian to determine the magnetic exchange couplings.
The spectra reveal signatures of Dirac points around the high symmetry point $X$ in the 2D Brillouin zone.   Our symmetry analysis and calculation based on linear spin wave theory shows that the Dirac point is due to the coexistence of easy-axis and easy-plane anisotropy, and is protected by symmetry.
On warming to temperatures above $T_N$, 
spin waves in the ordered state become paramagnetic scattering but still have short-range in-plane spin correlations that decay with increasing temperature. 
Our results therefore determine the magnetic exchange couplings of a zig-zag AF honeycomb lattice magnet, 
and provide a basis to compare with theoretical calculations using a Heisenberg-Kitaev Hamiltonian. 

\section{Experimental} 
BaNi$_2$(AsO$_4$)$_2$ crystallizes in a tetragonal structure with the in-plane and $c$ axis 
lattice parameters $a = 4.94$ \AA, $c = 23.43$ \AA, respectively (space group $R\bar{3}$) [Fig. 1(a)]. 
 Octahedrally coordinated Ni$^{2+}$ ions form a 2D honeycomb network, separated by AsO$_4$-Ba-AsO$_4$ layers [Fig. 1(b)]. 
The compound has an $ABC$ stacking along the $c$ axis. 
The distance between nearest neighbor Ni$^{2+}$ ions is 2.85 \AA, while the interlayer distance is 7.89 \AA, making 
an ideal 2D magnetic honeycomb lattice. 
Upon cooling, the system orders antiferromagnetically below $T_N\approx 18$ K with Ni$^{2+}$ spins forming 
zig-zag chains parallel or anti-parallel to the $a$ axis [Fig. 1(b)] \cite{Regnault1980}. 
Figure 1(c) shows a schematic of the reciprocal space in which the momentum transfer ${\bf Q} = H{\bf a^\ast} + K{\bf b^\ast} + L{\bf c^\ast}$ is denoted as $(H, K, L)$ r.l.u. The zig-zag AF transition at $T_N\approx 18$ K is confirmed by magnetic susceptibility [Fig. 1(d)] and neutron diffraction
measurements [Fig. 1(e)].  

The polycrystalline sample was synthesized using a solid state method. Stoichiometric powders of BaO (99.9 \%, Alfa Aesar), NiO (99.9 \%, Alfa Aesar) and As$_2$O$_5$ (99.9 \%, Alfa Aesar) were mixed, pressed and sintered at 850 $^\circ$C in a evacuated quartz tube. 
Plate shape single crystals of BaNi$_2$(AsO$_4$)$_2$ was grown by using NaCl as the flux with a molar ratio of $1:50$. 
Green transparent crystals [Fig. 1(e) inset] with typical size of 1 x 1 x 0.2 mm and clear hexagonal edges were separated from the flux by dissolving in hot water. 

We coaligned more than one hundred pieces of single crystals with a total mass of $\sim$1.5-g in the [$\mathit{H}$, $\mathit{K}$, 0] scattering plane and carried out time-of-flight inelastic neutron scattering experiments using the ARCS Spectrometer, Spallation Neutron Source, Oak Ridge National Laboratory. We used incident neutron energy of $E_i=16$ meV with high resolution mode to do rotational scans at 6 K, 17 K, 20 K, 37 K and 80 K. 

\section{Results and Discussion} 
The left panels in Figs. 2(a,b,c,d) show measured ${\bf Q}$-$E$ spectra of spin waves 
along high symmetry directions within the $[H,K]$ plane at $T = 6$ K.
The left panels in Figs. 2(e,f,g,h) are  ${\bf Q}$-dependence of spin waves in the $[H,K]$ plane
at energies $E=2\pm0.5, 3\pm 0.5, 4\pm 0.5, 6\pm0.5$ meV, respectively. 
The measured intensity below the line plots (white lines) between the $K$ and $\Gamma$ points in Fig. 2(a) and 2(d) are due to 
the finite integration range perpendicular to the cutting direction ($-0.1 \leq K_\perp \leq 0.1$). 
The spin wave spectra consist of four branches, consistent with the zig-zag AF structure 
that has four Ni$^{2+}$ ions in the magnetic unit cell within the 2D plane. 
For magnetic ordered systems, 
the ordered moment direction is typically determined by dipolar interactions or single-ion anisotropy associated with spin-orbit coupling (SOC), both can induce a spin gap at the ordering wave vector \cite{Lado}.
The size of the spin gap indicates the finite energy cost for the spins to fluctuate away from the ordered direction. 
The presence of the spin gaps at the $M$ points [Figs. 2(a-c)], 
especially the double-gap modes in the Figure 2(b) and 2(c), 
reveals that both easy-plane and easy-axis anisotropies should be taken into account to understand spin waves of BaNi$_2$(AsO$_4$)$_2$.

In order to determine the magnetic exchange couplings in BaNi$_2$(AsO$_4$)$_2$, we compared the measured spin waves with the spectra calculated using the linear spin wave theory via the SpinW program \cite{Toth}. 
We assumed a general Hamiltonian

\begin{align}\label{eq1}
 H = \sum_{i<j} (J_1 {\bf S}_i \cdot {\bf S}_j + J_2 {\bf S}_i \cdot {\bf S}_j + J_3 {\bf S}_i \cdot {\bf S}_j) \nonumber\\ 
 +\sum_j D_z (S_j^z)^2 + \sum_j D_x (S_j^x)^2,
\end{align}

where $J_1$, $J_2$ and $J_3$ are nearest, next nearest and next-next-nearest neighbour coupling constant [Fig. 1(b)], and the $D_z$ and $D_x$ terms in the equation 
describe the easy-plane and easy-axis anisotropies of the Ni$^{2+}$ ions, respectively.
The overall Hamiltonian is similar to the ones used to describe the similar compounds BaNi$_2$(PO$_4$)$_2$ and BaNi$_2$(VO$_4$)$_2$, where single-ion anisotropies are introduced \cite{Regnault1980,Regnault1986}.  We do not consider in-plane 
anisotropic interactions (Kitaev term) due to the small SOC for Ni$^{2+}$ ions 
in an octahedral local configuration \cite{Jackeli09,Takagi2019}. Interlayer magnetic exchange couplings were also not 
included because there were no evidence of spin wave modulation along the $L$ direction, suggesting negligible $c$-axis magnetic exchange coupling.

Since BaNi$_2$(AsO$_4$)$_2$ has zig-zag AF order, spin waves stem from the $M$ point where there is a 
strong magnetic Bragg peak [Figs. 2(b,c)].  To fit the spin wave spectra, we set the signs of initial magnetic exchange interactions to be consistent 
with the zig-zag magnetic order. Then the exchange parameters were varied manually to compare the simulated spectra with the experimental data. The best fit yields $J_1 = -0.69$, $J_2 = -0.03$, $J_3 = 1.51$, $D_z = 0.15$, and $D_x = -0.12$ meV [The right panels in Figs. 2(a-h)]. These results are compatible with the susceptibility measurement [Fig. 1(d)] through the relation $\Theta_{CW} = -S(S+1)(g-1)^2\Sigma_{i=n,nn,nnn} z_i J_i / 3k_B$, 
where $S = 1$ is the spin value, $k_B$ is the Boltzmann constant, $g$ is the Land\'e factor, and $z_i$ is the number of neighbors coupled to each magnetic ion by the magnetic exchange $J_i$ \cite{Klyushina2017, Choi2012}.
We note that the exchange couplings extracted from our measured spin waves are different from previous reports \cite{Regnault1980}. 
In spin wave dispersion along the $[H, H, 0]$, $[H, 0, 0]$, $[H - 0.5, H + 0.5, 0]$, and $[H -1/6, 1/3, 0]$ directions [Figs. 2(a-d)], 
and constant-energy cuts with different energies [Figs. 2(e-h)], the calculations describe the measured data quite well.
In particular, the anisotropy terms $D_z$ and $D_x$ reproduced the observed double-gap modes in the dispersion along the $[H, 0, 0]$, and $[H - 0.5,H + 0.5, 0]$ directions [Figs. 2(b) and 2(c)].

\begin{figure}[t]
\centering
\includegraphics[scale=.8]{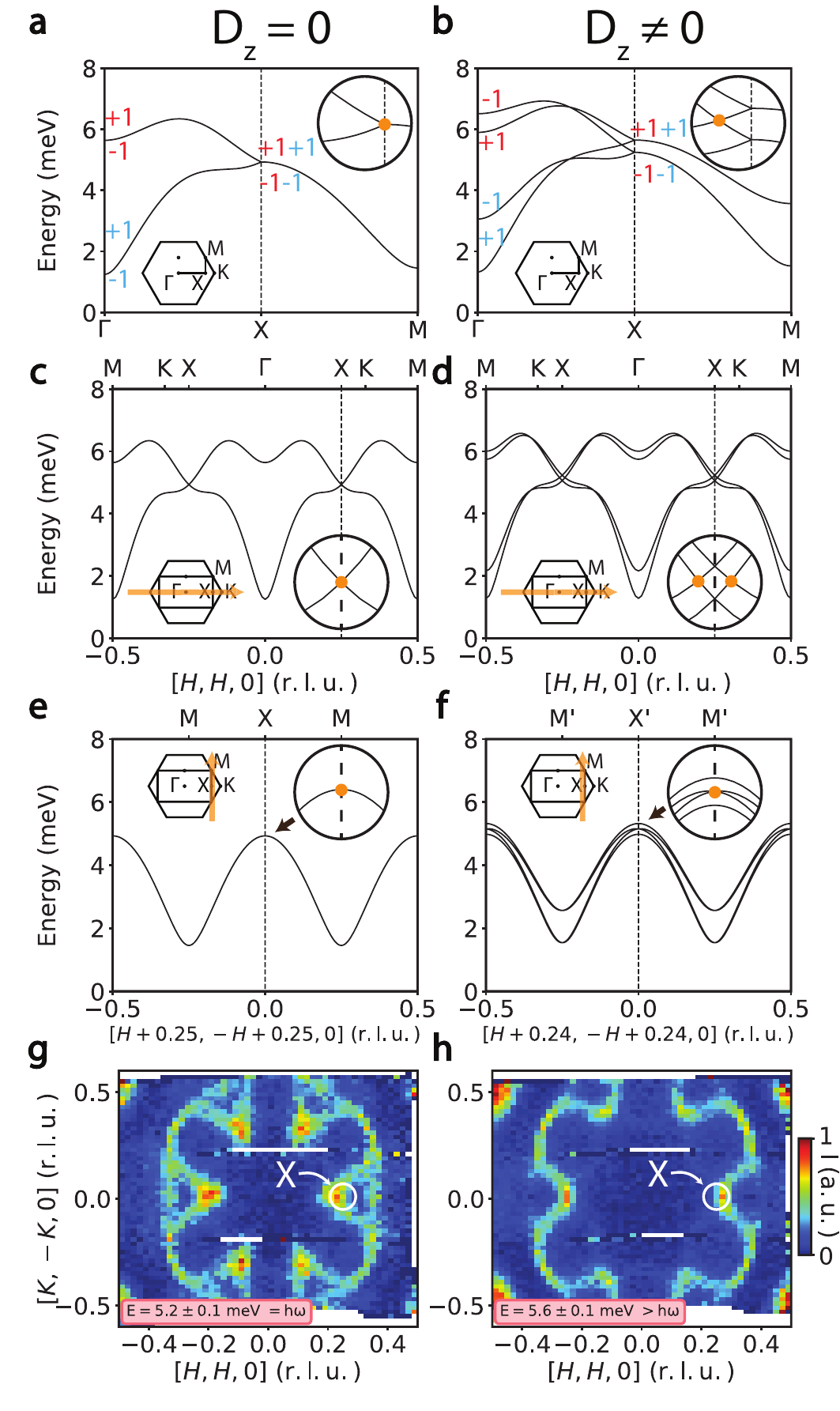}
\caption{\label{fig3} (a,b) Line plots of spin wave dispersion along high-symmetry directions with $D_z = 0$ (a) and 0.5 meV (b). The signs of the eigenvalues were labeled.  
(c,d) Calculated line plot of spin wave dispersion (not twinned) along the [$\mathit{H}$, $\mathit{H}$, 0] direction with $D_z = 0$ (c) and 0.2 meV (d).
(e,f) Calculated line plot of spin wave dispersion (not twinned) along perpendicular direction across the Dirac points in (c) and (d).
(g,h) Constant-energy cuts at selected energy-transfer in the $[H, K]$ plane at 6 K. 
The experimental data were integrated over $-4 \leq L \leq 4$.
}
\end{figure}

The magnetic interaction of BaNi$_2$(AsO$_4$)$_2$ can be described by the Hamiltonian in Equation \ref{eq1}. From the fitting above, we got
 $D_z > 0$ and
$D_x < 0$, describing the easy-plane and easy-axis anisotropies of the Ni$^{2 +}$
ions, respectively. Here, from symmetry arguments, we interpret the magnon
dispersion of the above Hamiltonian, and establish the existence of
topological Dirac points and nodal lines.

With the zigzag AF ordering, the system respects $M_x$, $M_y$,
$\bar{C}_{2 y} = t (\bm{a}_2) C_{2 y}$, $\bar{M}_z = t (\bm{a}_2)
M_z$ and $T \bar{M}_y = Tt (\bm{a}_1 / 2) M_y$ [Fig. 3(a)]. Here,
$\bm{a}_1 = (1, 0)$ and $\bm{a}_2 = \left( 1 / 2, \sqrt{3} / 2
\right)$ are the two lattice vectors (without the magnetic order). We will
start the discussion from a simplified $D_z = 0$ case, where the
$\mathcal{S}_x = \sum_i S^x_i$ is a conserved quantity. The magnon bands with
$D_z = 0$ is presented in Figs. 3(a). Along the $\mathrm{X}-\mathrm{M}$ line,
every band is fourfold degenerate, which splits into two doubly degenerate
bands on the $\Gamma -\mathrm{X}$ line. On the $\mathrm{X}-\mathrm{M}$ line,
$(T \bar{M}_y)^2 = - 1$, indicating a Kramers degeneracy. On the other hand,
$T \bar{M}_y$ commutes with $\mathcal{S}_x$, indicating the Kramers pair is
within the same eigenspace of $\mathcal{S}_x$. In addition, $\bar{C}_{2 y}$
anticommutes with $\mathcal{S}_x$, forcing different eigenspaces of
$\mathcal{S}_x$ to be degenerate, leading to the observed four-fold
degeneracy. The doubly degeneracy on the $\Gamma -\mathrm{X}$ line is
protected by the anticommuting operators $M_y S_x = - S_x M_y$ or $\bar{M}_z
S_x = - S_x \bar{M}_z$, meaning the two magnons in the doubly-degenerate band
belongs to separate eigenspaces of $M_y$ or $\bar{M}_z$.

For finite $D_z$, $\mathcal{S}_x$ is no longer a good quantum number.
Therefore, the doubly degeneracy on the $\Gamma -\mathrm{X}$ line is lifted.
Especially, the crossings between the originally two degenerate bands [Figs. 3(b)] 
are protected by $M_y$ or $\bar{M}_z$. These crossings are actually
nodal lines since $\bar{M}_z$ is respected in the whole Brillouin zone. Besides
the crosssings within the originally degenerate bands, there is another
crossing circled in red, which we will argue is inevitable for small $D_z$.
Especially, we will argue that the four degenerate bands (with $D_z = 0$) at
$\mathrm{X}$ are splitted into two degenerate states with $M_y$ eigenvalues
$\{ + 1, + 1 \}$ and $\{ - 1, - 1 \}$ [Fig. 3(c,d,e,f)]. To show this, we need to use the
following three commutation relations that holds at $\mathrm{X}$: $M_x M_y =
M_y M_x$, $M_y \bar{C}_{2 y} = \bar{C}_{2 y} M_y$ and $M_x \bar{C}_{2 y} = -
\bar{C}_{2 y} M_x$. We now use proof by contradiction and assume every doubly
degenerate band at $\mathrm{X}$ point has states $\{ \varphi_{+ 1}, \varphi_{-
1} \}$ with $M_y$ eigenvalues $\{ + 1, - 1 \}$. Since $M_x$ commutes with
$M_y$, $M_x \varphi_{+ 1}$ should also be a $M_y$ eigenstate with +1
eigenvalue. Therefore, $M_x \varphi_{+ 1} \propto \varphi_{+ 1}$, which means
$\varphi_{+ 1}$ is also an eigenstate of $M_x$ with eigenvalue $\lambda$.
Furthermore, since $M_x$ anticommutes with $\bar{C}_{2 y}$, $\bar{C}_{2 y}
\varphi_{+ 1}$ should be an eigenstate of $M_x$ with eigenvalue $- \lambda$.
As a result, $\bar{C}_{2 y} \varphi_{+ 1} \propto \varphi_{- 1}$. Finally, \
$M_y$ commutes with $\bar{C}_{2 y}$, indicating $\overline{C}_{2 y} \varphi_{+
1} \propto \varphi_{- 1}$ is an eigenstate of $M_y$ with eigenvalue $+ 1$.
This contradicts the original assumption that $\varphi_{- 1}$ is an eigenstate
of $M_y$ with eigenvalue $- 1$. Therefore, every doubly degenerate band at
$\mathrm{X}$ has $M_y$ eigenvalues $\{ + 1, + 1 \}$ or $\{ - 1, - 1 \}$. With
the knowledge of the $M_y$ eigenvalues on the $\Gamma -\mathrm{X}$ line and at
the $\mathrm{X}$ points, it is evident that the crossing circled in red is
unavoidable, which is a topological Dirac point [Figs. 3(c-f)].

To compare the above discussion with data, we show constant-energy cuts at [Fig. 3(g)] and slightly above [Fig. 3(h)] the energy $E = 5.2$ meV where we expect Dirac crossings. Although twinned domains and finite energy resolution give rise to complicated features in the constant-energy cuts, we can clearly see the maximums of intensity are around $X$ points as marked in Figs. 3(g) and 3(h),
 which provides direct evidence of the existence of the Dirac points in the spectrum. 

In strongly correlated electron materials such as the parent compound of iron based superconductors, in-plane spin-spin correlation are weakly dependent on the AF ordering temperature of the system \cite{Harriger2012}.  To test if this is also the case for 
zig-zag ordered BaNi$_2$(AsO$_4$)$_2$, we show in Fig. 4 temperature-dependence of the constant-energy cuts at $T=T_N-1.5=17$ K and $T=T_N+18.5=37$ K with energies of $E=0, 2, 3, 4, 5, 6$ meV. In the elastic channel,
magnetic ordered peaks at $M$ points below $T_N$ disappear at 37 K [Fig. 4(a)]. 
However, with finite energy transfer [Figs. 4(b-f)], although intensities dramatically drop upon warming up, the spin correlations persist up to 37 K.

\begin{figure}[t]
\centering
\includegraphics[scale=.8]{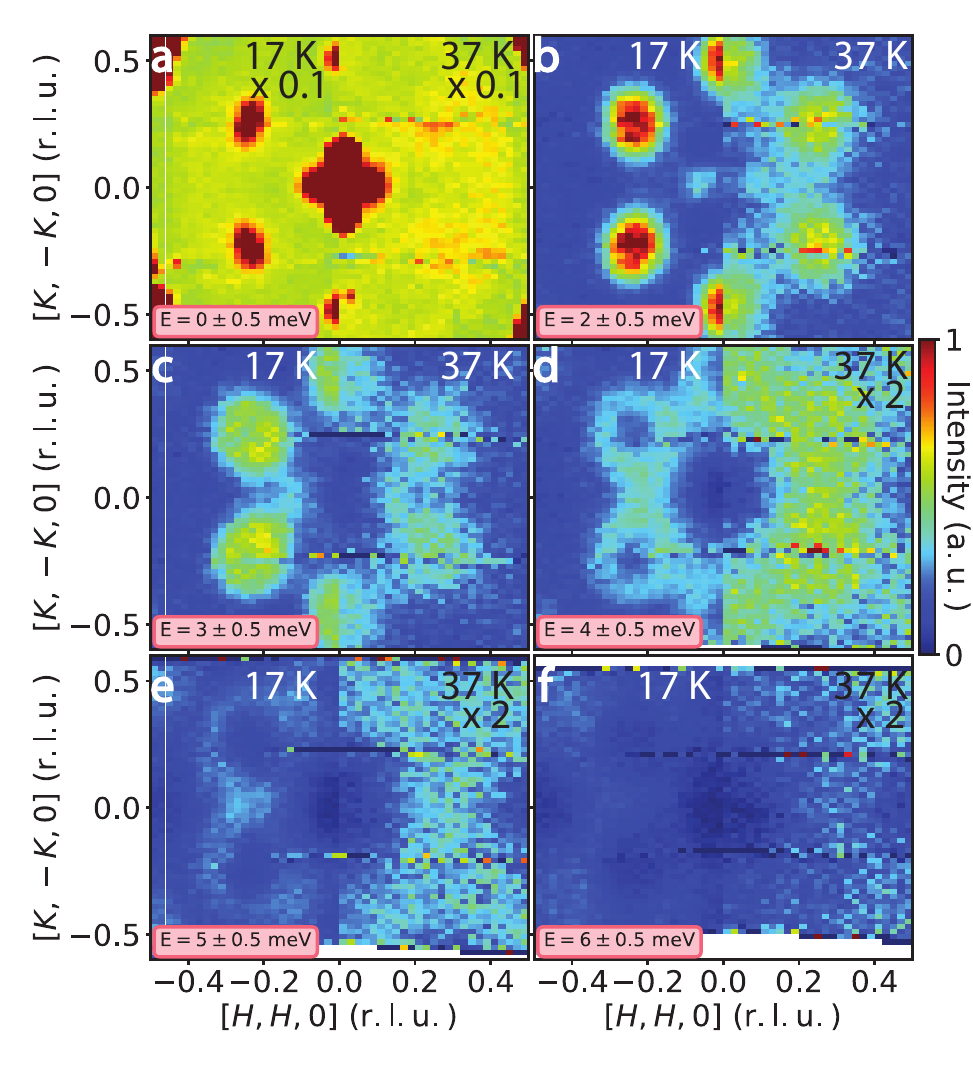}
\caption{(a-e) Temperature-dependence of constant-energy cuts at selected energy-transfer in the $[H, K]$ plane. The data were integrated over $-4 \leq L \leq 4$. The elastic scattering in the panel (a) and the inelastic scattering at 37 K showed in the right panels of (d-f) were multiplied by 0.1 and 2, respectively. We see clear spin correlations at $T=37 \approx 2 T_N$ K.
}
\end{figure}

By normalizing the observed magnetic scattering $S({\bf Q},E)$ with a vanadium standard, we can estimate 
the imaginary part of the dynamic susceptibility $\chi^{\prime\prime}({\bf Q},E)$ in absolute units via  
$\chi^{\prime\prime}({\bf Q}, E)=[1-\exp(-E/k_B T)] S({\bf Q}, E) $, 
where $k_B$ is the Boltzmann's constant.  The data points in Fig. 5(a) show the estimated energy dependence of the 
local dynamic susceptibility $\chi^{\prime\prime}(E)=\int\chi^{\prime\prime}({\bf Q}, E)d{\bf Q}/\int d{\bf Q}$, where the integration is within 
the first Brillouin zone \cite{Dai2015}.  Calculations using a Heisenberg Hamiltonian assuming $S=1$ shown in 
dashed and solid lines are comparable to the observation, thus confirming that $S=1$ Heisenberg Hamiltonian can describe the data. 
In correlated electron materials such as the parent compound of iron based superconductors, 
in-plane spin-spin correlation are weakly dependent on the AF ordering temperature of the system \cite{Harriger2012}.  
To test if this is also the case for zig-zag ordered BaNi$_2$(AsO$_4$)$_2$, we show in Fig. 5(b-d) 
temperature-dependence of the spin wave dispersion along the $[H, 0, 0]$ direction.
In the elastic channel,
magnetic ordered peaks at $M$ points disappear above $T_N$ [Fig. 1(e)]. 
However, with finite energy transfer [Figs. 5(b-d)], the in-plane spin-spin correlations persist up to even 80 K, consistent with the 2D nature of the magnetic scattering \cite{Wicksted}. 

\begin{figure}[t]
\centering
\includegraphics[scale=.8]{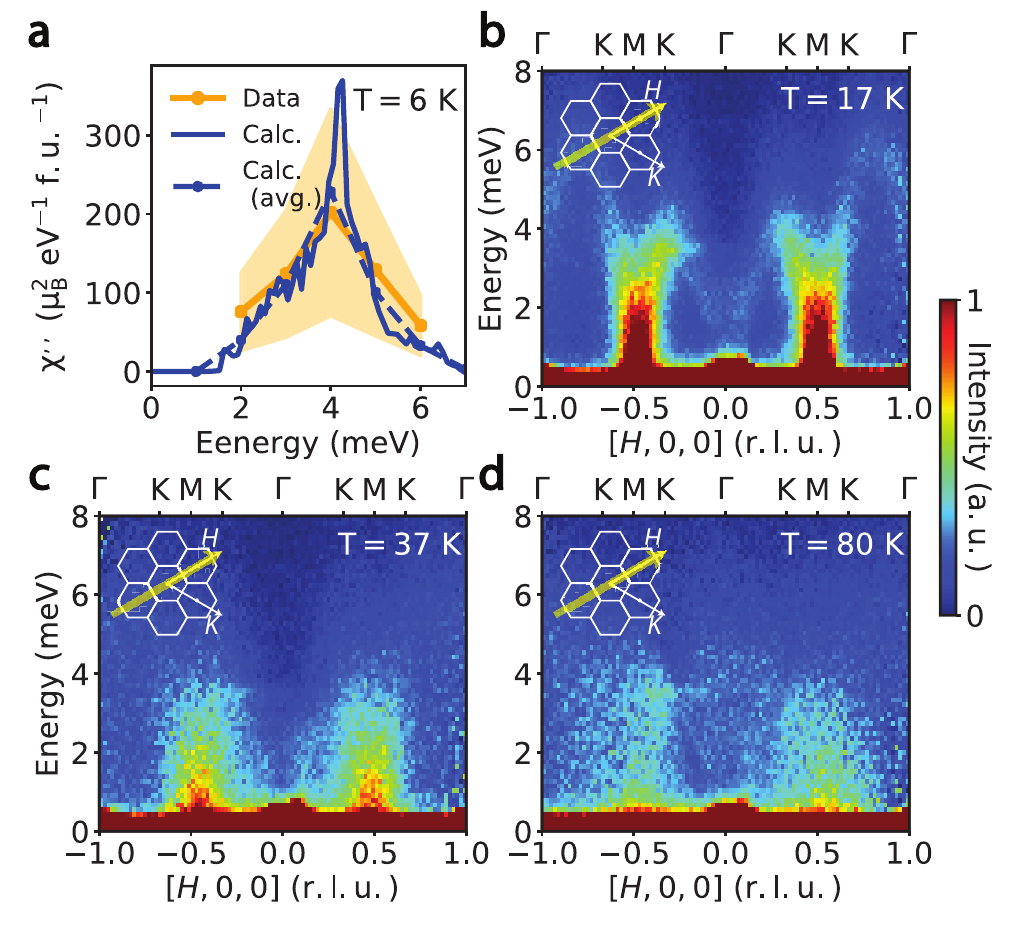}
\caption{\label{fig4} (a) Measured (orange) and calculated (blue) Energy-dependent local dynamic susceptibility of BaNi$_2$(AsO$_4$)$_2$, obtained by integration in the first Brillouin zone in the $[H, K]$ plane. The light orange region indicates the uncertainty of the experiment data mainly from the normalization process. Calculated local dynamic susceptibility averaged every meV is also plotted to directly compare with measured data.
(b-d) Temperature dependence of neutron scattering $E$-${\bf Q}$ spectra along 
the $[H, 0, 0]$ direction observed at 17 K, 37 K, and 80 K. The experimental data were integrated over $-4 \leq L \leq 4$ and 
$-0.1 \leq K_\perp \leq 0.1$. 
}
\end{figure}

To compare magnetic exchange couplings of the zig-zag ordered BaNi$_2$(AsO$_4$)$_2$ 
with related compounds BaNi$_2$(PO$_4$)$_2$ and BaNi$_2$(VO$_4$)$_2$, we note 
that the latter compounds have different magnetic structures. 
The local moments on each Ni site in BaNi$_2$(PO$_4$)$_2$ and BaNi$_2$(VO$_4$)$_2$ are anti-parallel to each other, and the direction of the moments in BaNi$_2$(PO$_4$)$_2$ is along the bond. 
Table I summarizes the nearest, second-nearest, third-nearest magnetic exchange couplings, easy-plane and easy-axis anisotropy, and Kitaev and $\Gamma$ terms in BaNi$_2$(XO$_4$)$_2$ (X = V, P, As) and $\alpha$-RuCl$_3$.

\begin{table}[]
\centering
\caption{
The estimated magnetic interactions and anisotropies in BaNi$_2$(XO$_4$)$_2$ (X = V, P, As) \cite{Regnault1980,Regnault1986} and $\alpha$-RuCl$_3$ \cite{Kim2016,Wu2018} from representative studies. 
All units are converted to meV.
}
\begin{tabular}{c|c|c|c|c|c}
\hline
 & BaNi$_2$As$_2$O$_8$ & BaNi$_2$V$_2$O$_8$    & BaNi$_2$P$_2$O$_8$  	& $\alpha$-RuCl$_3$	 & $\alpha$-RuCl$_3$  \cr
\hline
$J_1$ 		& -0.69		& 12.3		& 0.19		& -1.53		& -0.35			\cr
$J_2$ 		& -0.03		& 1.25		& 0.025 	&			&			\cr
$J_3$ 		&  1.51		& 0.2		& 0.76		& 			& 0.34		\cr
$D_z$ 		&  0.15		& 0.07		& 0.269		&			&			\cr
$D_x$		& -0.12		&-0.001		&			&			&			\cr
$K$ 		&			&			&			& -6.55		& 5.25		\cr
$\Gamma$&			&			&			& -2.8		& 2.4		\cr

\hline
\end{tabular}
\label{tab:my_label}
\end{table}

As we can see from the table, the magnetic structure and properties are very sensitive to exchange couplings in an AF honeycomb lattice magnet. 
In previous neutron scattering work on single crystals of BaM$_2$(XO$_4$)$_2$ (M = Ni, Co; X = V, P, As), spin wave spectra were only measured along selected high symmetry directions with poor resolution using a triple-axis neutron spectrometer \cite{Regnault83,Regnault1986,Regnault1983,Regnault1990,Regnault2018}, much different from the entire spin wave spectra obtained using a high resolution neutron time-of-flight chopper spectrometer (Figs. 2-4) \cite{ARCS}. 
The $J_1$, $J_2$, and $J_3$ 
parameters determined in our experiment for BaNi$_2$(AsO$_4$)$_2$
are consistent with the proposed phase diagram for a $S=1$ zig-zag ordered antiferromagnet
in a classical Heisenberg Hamiltonian \cite{Boyko2018,Fouet2001}. 

The evolution of magnetic Dirac bosons in the honeycomb lattice has been calculated before in two papers. Several AF structures have been discussed, including simple AF, zig-zag, dimerized, armchair and stripe. In the simple AF structure, the breaking of inversion symmetry eliminates the presence of a Dirac point, while in more complex AF configurations, the crossing of spin waves can produce Dirac-like nodes. However, there is no experimental confirmation from any real materials of complex AF structure. In the case of CoTiO$_3$, it is actually ferromagnetic correlation in the 2D honeycomb lattice. Also notice that the Dirac points in BaNi$_2$(AsO$_4$)$_2$ is not on the boundary of Brillouin zone, not like the case in 2D FM honeycomb lattice, CoTiO$_3$ nor the 3D antiferromagnet Cu$_3$TeO$_6$ \cite{Yao2018,Bao2018}. This is due to the more generalized magnetic structure in BaNi$_2$(AsO$_4$)$_2$, which has lower symmetry.

\section{Conclusions}
In summary, we used time-of-flight inelastic neutron scattering to study spin waves of the $S=1$ honeycomb lattice antiferromagnet BaNi$_2$(AsO$_4$)$_2$, which has a zig-zag AF ground state.  We determine the magnetic exchange interactions 
in the zig-zag AF ordered phase, and show that spin waves in BaNi$_2$(AsO$_4$)$_2$ have 
symmetry-protected Dirac points inside the Brillouin zone boundary. These results provide a microscopic understanding of the 
zig-zag AF order and associated Dirac magnons in honeycomb lattice magnets, and are also important for establishing the magnetic interactions in Kitaev quantum spin liquid candidates.

\section{acknowledgments}
The neutron scattering and synthesis work at Rice is supported by the US NSF DMR-2100741 and 
the Robert A. Welch Foundation Grant No. C-1839, respectively (P.D.). 
	C.W. and D.X. acknowledge the support of  AFOSR MURI 2D MAGIC (FA9550-19-1- 0390).
	A portion of this research used
	resources at the Spallation Neutron Source, 
	a DOE Office of Science User Facilities operated
	by ORNL.

\section{appendix}
Here we present detailed symmetry analysis of the
magnon band structure. In the main text, we have argued that the bands on the
$\Gamma$-X are doubly degenerate with $\{ + 1, - 1 \}$ $M_y$ eigenvalues for
$D_z = 0$. A finite $D_z$ splits the degeneracy. Here, we show that the doubly
degenerate bands at the X point have $M_y$ eigenvalues $\{ + 1, + 1 \}$ or $\{
- 1, - 1 \}$ for finite $D_z$.

The relevant symmetries at the X point are $M_y$, $\overline{C}_{2 y} \assign
t (\tmmathbf{a}_2) C_{2 y}$ (the combination of two-fold rotation symmetry and
a translation along $\tmmathbf{a}_2$) and $M_x$. Here, $\tmmathbf{a}_1 = (1,
0)$ and $\tmmathbf{a}_2 = \left( 1 / 2, \sqrt{3} / 2 \right)$ are the lattice
vectors of the nonmagnetic unit cell [see Fig. 1(b) in the main text]. We now
work out the commutation relation between the symmetry operators:
\begin{eqnarray*}
  M_x M_y & = & M_y M_x ;
\end{eqnarray*}
\begin{eqnarray*}
  M_y \overline{C}_{2 y} & = & M_y t (\tmmathbf{a}_2) C_{2 y}\\
  & = & t (-\tmmathbf{a}_2 +\tmmathbf{a}_1) M_y C_{2 y}\\
  & = & t (- 2\tmmathbf{a}_2 +\tmmathbf{a}_1) \overline{C}_{2 y} M_y\\
  & = & \overline{C}_{2 y} M_y ;
\end{eqnarray*}
\begin{eqnarray*}
  M_x \overline{C}_{2 y} & = & M_x t (\tmmathbf{a}_2) C_{2 y}\\
  & = & t (-\tmmathbf{a}_1 +\tmmathbf{a}_2) M_x C_{2 y}\\
  & = & t (-\tmmathbf{a}_1) \overline{C}_{2 y} M_x\\
  & = & - \overline{C}_{2 y} M_x ;
\end{eqnarray*}
where we have used the fact that $t (- 2\tmmathbf{a}_2 +\tmmathbf{a}_1) = 1$
and $t (-\tmmathbf{a}_1) = - 1$ at X point. We now use proof by contradition
and assume one doubly denerate band at X point contains states $\{ \varphi_{+ 1},
\varphi_{- 1} \}$ with $M_y$ eigenvalues $\{ + 1, - 1 \}$. Since $M_x$
commutes with $M_y$, $M_x \varphi_{+ 1}$ should also be a $M_y$ eigenstate
with +1 eigenvalue. Therefore, $M_x \varphi_{+ 1} \propto \varphi_{+ 1}$,
which means $\varphi_{+ 1}$ is also an eigenstate of $M_x$ with eigenvalue
$\lambda$. Furthermore, since $M_x$ anticommutes with $\overline{C}_{2 y}$,
$\overline{C}_{2 y} \varphi_{+ 1}$ should be an eigenstate of $M_x$ with
eigenvalue $- \lambda$. As a result, $\overline{C}_{2 y} \varphi_{+ 1} \propto
\varphi_{- 1}$. Finally, \ $M_y$ commutes with $\overline{C}_{2 y}$,
indicating $\overline{C}_{2 y} \varphi_{+ 1} \propto \varphi_{- 1}$ is an
eigenstate of $M_y$ with eigenvalue $+ 1$. This contradicts the original
assumption that $\varphi_{- 1}$ is an eigenstate of $M_y$ with eigenvalue $-
1$. Therefore, every doubly degenerate band at X has $M_y$ eigenvalues $\{ +
1, + 1 \}$ or $\{ - 1, - 1 \}$.

{}
\end{document}